\documentclass[a4paper,11pt]{article}
\usepackage{pos}

\title{Topology in high-$T$ QCD via staggered spectral projectors}

\author[a,b]{Andreas Athenodorou}
\author*[c]{Claudio Bonanno}
\author[a]{Claudio Bonati}
\author[d]{Giuseppe Clemente}
\author[a]{Francesco D'Angelo}
\author[a]{Massimo D'Elia}
\author[a]{Lorenzo Maio}
\author[e]{Guido Martinelli}
\author[f]{Francesco Sanfilippo}
\author[g,h,i]{Antonino Todaro}

\affiliation[a]{Università di Pisa and INFN Sezione di Pisa, Largo B.~Pontecorvo 3, I-56127 Pisa, Italy}

\affiliation[b]{Computation-based Science and Technology Research Center, The Cyprus Institute, 20 Kavafi Str., Nicosia 2121, Cyprus}

\affiliation[c]{Centro Nazionale INFN di Studi Avanzati ``The Galileo Galilei Institute for Theoretical Physics'' (GGI), Largo E.~Fermi, 2, I-50125, Arcetri, Florence, Italy}

\affiliation[d]{Institute for Mathematics, Astrophysics and Particle Physics (IMAPP), Radboud University Nijmegen, Heyendaalseweg 135, 6525 AJ Nijmegen, The Netherlands}

\affiliation[e]{Dipartimento di Fisica and INFN Sezione di Roma ``La Sapienza'', Piazzale Aldo Moro 5, I-00185 Rome, Italy}

\affiliation[f]{INFN Sezione di Roma Tre, Via della Vasca Navale 84, I-00146 Rome, Italy}

\affiliation[g]{Department of Physics, University of Cyprus, P.O.~Box 20537, 1678 Nicosia, Cyprus}

\affiliation[h]{Faculty of Mathematics und Natural Sciences, University of Wuppertal, Wuppertal-42119, Germany}

\affiliation[i]{Dipartimento di Fisica, Università di Roma ``Tor Vergata'', Via della Ricerca Scientifica 1, I-00133 Rome, Italy}

\emailAdd{a.athenodorou@cyi.ac.cy}
\emailAdd{claudio.bonanno@pi.infn.it}
\emailAdd{claudio.bonati@unipi.it}
\emailAdd{giuseppe.clemente@ru.nl}
\emailAdd{francesco.dangelo@phd.unipi.it}
\emailAdd{massimo.delia@unipi.it}
\emailAdd{lorenzo.maio@phd.unipi.it}
\emailAdd{guido.martinelli@roma1.infn.it}
\emailAdd{francesco.sanfilippo@infn.it}
\emailAdd{atodar01@ucy.ac.cy}

\abstract{We present preliminary lattice results for the topological susceptibility in high-$T$ $N_f=2+1$ QCD obtained discretizing this observable via spectral projectors on eigenmodes of the staggered operator, and we compare them with those obtained with the standard gluonic definition. The adoption of the spectral discretization is motivated by the large lattice artifacts affecting the continuum scaling of the gluonic susceptibility at high $T$, related to the choice of non-chiral fermions in the action.}

\FullConference{The 38$^{\text{th}}$ International Symposium on Lattice Field Theory, LATTICE2021 26$^{\text{th}}$-30$^{\text{th}}$ July, 2021 Zoom/Gather@Massachusetts Institute of Technology}

\usepackage{braket}
\usepackage[shortlabels]{enumitem}
\usepackage{placeins}

\DeclareMathOperator{\Tr}{\mathrm{Tr}}
\DeclareMathOperator{\round}{\mathrm{round}}

\newcommand{\beq}{\begin{eqnarray}}
\newcommand{\eeq}{\end{eqnarray}}
\newcommand{\YM}{\mathit{YM}}
\newcommand{\stag}{\mathit{stag}}
\newcommand{\SP}{\mathit{SP}}

\begin{document}
\maketitle

\section{Introduction}
In recent years there has been a renewed interest in the study of the topological properties of QCD at high temperatures because of their phenomenological relevance in the context of axion cosmology~\cite{Bonati:2015vqz, Petreczky:2016vrs, Frison:2016vuc, Borsanyi:2016ksw}. As a matter of fact, the axion effective (i.e., temperature-dependent) physical parameters are related to QCD topological observables, such as the topological susceptibility $\chi = \braket{Q^2}/V$, where
\beq\label{eq:topocharge_continuum}
Q=\frac{1}{32 \pi^2} \varepsilon_{\mu\nu\rho\sigma}\int d^4x \Tr\{F^{\mu\nu}(x)F^{\rho \sigma}(x)\}
\eeq
is the topological charge and $V$ the $4$-dimensional space-time volume.

At asymptotically-high temperatures one expects the leading-order perturbative picture to become more and more reliable, allowing to obtain analytic control on $\chi$ as a function of $T$ via the Dilute Instanton Gas Approximation (DIGA), which predicts $\chi(T) \sim T^{-\alpha}$, with $\alpha\simeq 8$ for 3 light quarks~\cite{Gross:1980br}. At temperatures of the order of 1~GeV or below, non-perturbative effects are instead more important, and the lattice approach is a fundamental tool to obtain a non-perturbative determination of $\chi(T)$ from first principles at these energy scales.

The study of topological properties of QCD from the lattice is however extremely challenging because of many non-trivial computational problems affecting high-temperature simulations, the toughest ones being:
\begin{enumerate}[label=(\textit{\roman*})]
\item\label{enum:prob_1} \emph{Dominance of the $Q=0$ sector}\\
The susceptibility is computed from the fluctuations of the topological charge. Being $\chi$ suppressed at high $T$, $\braket{Q^2} = \chi V \ll 1$ on affordable volumes, meaning that fluctuations of $Q$ become extremely rare. Thus, unfeasible large statistics is required to obtain a meaningful estimation of $\chi$.
\item\label{enum:prob_2} \emph{Chiral symmetry and large lattice artifacts}\\
Due to the presence of the determinant of the Dirac operator, the contribution to the path integral of configurations with non-zero topological charge is suppressed by powers of the quark masses. This aspect is responsible for the vanishing of $\chi$ in the chiral limit. On the lattice, however, this suppression is not as efficient as in the continuum if non-chiral fermions (e.g., staggered) are adopted to discretize the action, being exact zero-modes absent from the spectrum of the lattice quark matrix in this case. This results in large lattice corrections that affect the continuum scaling of $\chi$, and makes it harder to have systematics related to the continuum extrapolation under control.
\item\label{enum:prob_3} \emph{Topological freezing}\\
As the lattice spacing $a$ is reduced and the continuum limit is approached, the Monte Carlo Markov chain generated from standard algorithms tends to remain trapped in a fixed topological sector~\cite{Vicari:2008jw, Schaefer:2010hu, Bonati:2017woi}. More precisely, the number of updating steps necessary to change the topological charge of a configurations diverges exponentially as a function of $a^{-1}$, making simulations close to the continuum limit extremely challenging.
\end{enumerate}

Some strategies have been proposed in the literature to overcome such problems, however, they rely on non-trivial assumptions. For instance, in Ref.~\cite{Borsanyi:2016ksw}, issues~\ref{enum:prob_1} and~\ref{enum:prob_3} were bypassed by giving up the sampling of the full distribution of the topological charge and focusing on just the sectors with $\vert Q \vert=0$ and 1. This approach can be justified assuming DIGA. Indeed, if topological objects are exactly non-interacting, then the contribution of these sectors is sufficient to reconstruct the full distribution. Issue~\ref{enum:prob_2}, instead, was overcome in Ref.~\cite{Borsanyi:2016ksw} by an \emph{a posteriori} reweighting of configurations with $Q\ne 0$ with the corresponding lowest eigenvalues of the continuum Dirac operator, so that corrections to the continuum limit are reduced. This procedure, however, may introduce undesired systematics since it modifies the path integral distribution used to compute expectation values.

In this talk we present new results towards an independent determination of $\chi$ in full QCD at high-$T$ without any extra assumption. In particular, we adopt the Spectral Projectors (SP) method~\cite{Luscher:2004fu, Giusti:2008vb, Luscher:2010ik, Cichy:2015jra, Alexandrou:2017bzk} defined for staggered fermions in Ref.~\cite{Bonanno:2019xhg} to obtain a definition of the topological susceptibility which, while reducing to the correct definition in the continuum limit, suffers from smaller lattice artifacts compared to the standard gluonic discretization. This allows to overcome issue~\ref{enum:prob_2} without the need for extra assumptions, allowing to perform more controlled continuum extrapolations in the typical lattice spacing range usually employed in full QCD simulations, thus avoiding the need of smaller lattice spacings (which would require to fight issue~\ref{enum:prob_3}). Moreover, we combine this technique with the multicanonic algorithm already applied in Refs.~\cite{Jahn:2018dke, Bonati:2018blm}, which allows to overcome issue~\ref{enum:prob_1} by adding a bias potential to the original action that enhances the fluctuations of suppressed topological sectors. Path integral expectation values with respect to the original distribution are then exactly obtained from multicanonic simulations through the standard reweighting technique.

This work is organized as follows: in Sec.~\ref{sec:num_setup} we present our numerical setup, in Sec.~\ref{sec:results} we present continuum extrapolated results obtained for the topological susceptibility from staggered spectral projectors for a temperature $T\simeq 430$~MeV above the transition and finally in Sec.~\ref{sec:conclusions} we draw our conclusions and discuss future outlooks.

\section{Numerical setup}\label{sec:num_setup}
\subsection{Lattice action}
We discretize $N_f=2+1$ flavors QCD on a $N_t \times N_s^3$ lattice adopting rooted stout-smeared staggered fermions for the quark sector and the tree-level Symanzik-improved action for the gauge sector. Our partition function is:
\beq\label{eq:partition_function}
Z_{\mathit{LQCD}}^{(2+1)}=\int [dU] e^{-S^{(L)}_\YM[U]} \det\left\{\mathcal{M}^{(\stag)}_{l}[U]\right\}^{\frac{1}{2}} \det\left\{\mathcal{M}^{(\stag)}_{s}[U]\right\}^{\frac{1}{4}},
\eeq
where $u=d=l$ and $s$ are the light and strange quark flavors,
\begin{gather}
\mathcal{M}^{(\stag)}_f[U](x,y) \equiv D_{\stag}[U^{(2)}](x,y) + \hat{m}_f \delta_{x,y},\\
\hat{m}_f \equiv am_f, \quad x_\mu, y_\mu \in \text{Lattice},\nonumber\\
\label{eq:stag_operator}
D_{\stag}[U^{(2)}](x,y) = \sum_{\mu=1}^{4}\eta_{\mu}(x)\left( U^{(2)}_{\mu}(x) \delta_{x,y-\hat{\mu}} - {U_{\mu}^{(2)}}^{\dagger}(x-\hat{\mu}) \delta_{x,y+\hat{\mu}} \right),\\ \eta_{\mu}(x) = (-1)^{x_1+\dots+x_{\mu-1}}, \nonumber
\end{gather}
is the staggered fermion matrix built using gauge links after $n_{\mathit{stout}}=2$ levels of stout smearing~\cite{Morningstar:2003gk} with isotropic stouting parameter $\rho_{\mathit{stout}}=0.15$, and
\begin{gather}
S_{\YM}^{(L)}[U] = - \frac{\beta}{3} \sum_{x, \mu \ne \nu} \left\{ c_0\Re\Tr\left[\Pi_{\mu\nu}^{(1\times1)}(x)\right] + c_1\Re\Tr\left[\Pi_{\mu\nu}^{(1\times2)}(x)\right]\right\}, \\ \beta=\frac{6}{g^2}, \quad c_0=\frac{5}{6}, \quad  c_1=-\frac{1}{12}, \nonumber
\end{gather}
is the tree-level Symanzik-improved gauge action built using the $n\times m$ plaquettes $\Pi_{\mu\nu}^{(n\times m)}(x)$ starting from site $x$ on the plane $\mu$--$\nu$ and defined in the terms of the non-stouted links, which are the integration variables of path integral~\eqref{eq:partition_function}. The choice of stout smearing to define the lattice Dirac operator allows to sample path integral~\eqref{eq:partition_function} through the standard Rational Hybrid Monte Carlo (RHMC) algorithm~\cite{Clark:2006fx, Clark:2006wp}, as stouted links are differentiable with respect to the non-stouted ones. The bare parameters $\beta$, $\hat{m}_s$ and $\hat{m}_u=\hat{m}_d \equiv \hat{m}_l$ are chosen in order to move on a Line of Constant Physics (LCP) while approaching the continuum limit $\beta \to \infty$. Our LCP has $m_{\pi} \simeq 135$~MeV and $\hat{m}_s / \hat{m}_l = m_s/m_l \simeq 28.15$ fixed to the physical values~\cite{Aoki:2009sc, Borsanyi:2010cj, Borsanyi:2013bia}.

\subsection{Topological charge discretizations}
The standard gluonic discretization of the topological charge~\eqref{eq:topocharge_continuum} can be expressed in terms of the plaquettes $\Pi^{(n\times m)}_{\mu\nu}(x)$. The simplest parity-defined discretization is the \emph{clover} definition:
\beq\label{eq:clover_charge}
Q_{\mathit{clov}} = \frac{-1}{2^9 \pi^2}\sum_{x}\sum_{\mu\nu\rho\sigma=\pm1}^{\pm4}\varepsilon_{\mu\nu\rho\sigma}\Tr\left\{\Pi_{\mu\nu}^{(1\times1)}(x)\Pi_{\rho\sigma}^{(1\times1)}(x)\right\}.
\eeq
This definition renormalizes multiplicatively and additively: $Q = Z_Q Q_{\mathit{clov}} + \eta_Q$~\cite{Campostrini:1988cy, DiVecchia:1981aev, DiVecchia:1981hh}. To deal with these renormalizations, we compute $Q_{\mathit{clov}}$ on smoothed configurations, as smoothing algorithms damp ultra-violet fluctuations while leaving the topological content of the configuration unchanged. Several methods can be adopted (e.g., stout smearing~\cite{Morningstar:2003gk}, cooling~\cite{Berg:1981nw,Iwasaki:1983bv,Itoh:1984pr,Teper:1985rb,Ilgenfritz:1985dz,Campostrini:1989dh,Alles:2000sc}, gradient flow~\cite{Luscher:2009eq, Luscher:2010iy}), all agreeing when properly matched with each other~\cite{Alles:2000sc, Bonati:2014tqa, Alexandrou:2015yba}. Here we adopt cooling for its numerical cheapness. In all our simulations $n_{\mathit{cool}} \sim 80$ are sufficient to obtain stable determinations of $Q_{\mathit{clov}}$, and this number is quite independent of the lattice spacing. For this reason, we adopt $n_{\mathit{cool}} = 100$ in every case. Since $Q_{\mathit{clov}}^{(\mathit{cooled})}$ is not integer valued, we adopt the prescription of Ref.~\cite{DelDebbio:2002xa} to round it:
\beq
Q_{\mathit{gluo}} = \round\left\{\alpha Q_{\mathit{clov}}^{(\mathit{cooled})}\right\}, \quad \text{with} \quad \alpha=\min_{x \geq 1} \left\langle\left[x Q_{\mathit{clov}}^{(\mathit{cooled})} - \round\left\{x Q_{\mathit{clov}}^{(\mathit{cooled})}\right\}\right]^2\right\rangle,
\eeq
where $\alpha$ is chosen so that the peaks of the distribution of $Q_{\mathit{clov}}^{(\mathit{cooled})}$ are located around integer values. Since the smoothed gluonic charge has $Z_Q \simeq 1$ and $\eta_Q \simeq 0$, the gluonic susceptibility is then simply obtained via
\beq
\chi_{\mathit{gluo}} = \frac{\braket{Q^2_{\mathit{gluo}}}}{V}, \qquad V=a^4 N_t N_s^3.
\eeq

Since the gluonic definition suffers from large lattice artifacts in the presence of non-chiral lattice quarks in the action, we also consider the staggered SP definition derived in Ref.~\cite{Bonanno:2019xhg}. This definition is based on a discretized version of the \emph{index theorem}:
\beq
Q = \Tr\{\gamma_5\} \,\, \rightarrow \,\, Q_{\SP, 0}^{(\stag)} = 2^{-d/2} \Tr\left\{\mathbb{P}_M \Gamma_5\right\}, \qquad \qquad \Gamma_5 = \gamma_5^{(\stag)},
\eeq
where $\mathbb{P}_M$ is the spectral projector over eigenmodes $\vert\lambda\vert \leq aM$ of the very same staggered operator~\eqref{eq:stag_operator} used in the lattice action and where $2^{-d/2}$ accounts for taste degeneration (with $d$ the space-time dimension). Here, the sum over eigenvalues up to the threshold mass $M$ substitutes the sum over zero-modes that would be sufficient in the continuum, since no exact zero-mode is present in the staggered spectrum. This definition renormalizes only multiplicatively
\beq
Q_{\SP}^{(\stag)} = Z_{\SP}^{(\stag)} Q_{\SP,0}^{(\stag)}, \qquad Z_{\SP}^{(\stag)} = \sqrt{\frac{\braket{\Tr\left\{\mathbb{P}_M\right\}}}{\braket{\Tr\left\{\Gamma_5 \mathbb{P}_M \Gamma_5 \mathbb{P}_M\right\}}}},
\eeq
thus the SP expression of the topological susceptibility is simply:
\beq
\chi_{\SP}^{(\stag)} = \frac{\braket{{Q_{\SP}^{(\stag)}}^2}}{V} = {Z_{\SP}^{(\stag)}}^2 \frac{\braket{{Q_{\SP,0}^{(\stag)}}^2}}{V} =  2^{-d}\frac{\braket{\Tr\{\mathbb{P}_M\}}}{\braket{\Tr\{\Gamma_5 \mathbb{P}_M \Gamma_5 \mathbb{P}_M \}}} \frac{\braket{\Tr\{\Gamma_5 \mathbb{P}_M\}^2}}{V}.
\eeq
In our calculations, we expressed the spectral projector $\mathbb{P}_M$ in terms of the eigenvectors of operator~\eqref{eq:stag_operator}:
\beq
\mathbb{P}_M \equiv \sum_{\vert \lambda \vert \le aM} u_{\lambda} u_{\lambda}^{\dagger}, \qquad \qquad i D_{\stag}[U^{(2)}] u_{\lambda} = \lambda u_{\lambda}, \quad \lambda \in \mathbb{R}.
\eeq

The bare parameter $M$ can be freely chosen in order to tune the corrections to the continuum limit, since its value does not affect the continuum extrapolation (as long as all the relevant modes are included), as in the continuum limit only zero-modes contribute to $\chi$. The only prescription that must be adopted is that the physical value of the renormalized threshold $M_R$ has to be kept constant as the continuum limit is approached. Given that $M$ renormalizes as a quark mass and that a LCP is known, it is sufficient to keep the ratio $M/m_f$, where $m_f$ is the bare mass of a given flavor $f$, constant as the lattice spacing is varied in order to approach the continuum limit at fixed physical value of $M_R$. In the following we will express the threshold mass $M$ in terms of the strange quark mass $m_s$.

\subsection{Multicanonic algorithm}
The multicanonic approach consists in adding to the action a topological bias potential $V_{\mathit{topo}}(Q_{\mathit{MC}})$ tailored so that it enhances the probability of visiting suppressed topological sectors:
\beq
S_{\YM}^{(L)}[U] \rightarrow S_{\YM}^{(L)}[U] + V_{\mathit{topo}}(Q_{\mathit{MC}}).
\eeq

The quantity $Q_{\mathit{MC}}$ is a suitable discretization of the topological charge which does not need to be the same adopted for the measure of topological observables. It is important that $Q_{\mathit{MC}}$ is chosen sufficiently close to the physical topological charge so that the new path-integral distribution obtained adding the potential has a reasonable overlap with the starting one.

Following Ref.~\cite{Bonati:2018blm}, we choose:
\beq\label{eq:topo_potential}
V_{\mathit{topo}}(x)=
\begin{cases}
-\sqrt{ (Bx)^2 + C}, &\quad \vert x \vert < Q_{\mathit{max}},\\
-\sqrt{ (BQ_{\mathit{max}})^2 + C}, &\quad \vert x \vert \geq Q_{\mathit{max}}.
\end{cases}
\eeq
In our simulations, $Q_{\mathit{max}}=2$ or 3 proved to be enough to obtain enhancement of several orders of magnitude of suppressed sectors, allowing to observe lots of fluctuations of $Q_{\mathit{gluo}}$ during the Monte Carlo evolution. The values of $B$ and $C$ are tuned through short test runs in order to obtain reasonable Monte Carlo histories for the topological charge, where reasonable means that we are able to uniformly explore the interval $[-Q_{\mathit{max}},\,Q_{\mathit{max}}]$ without breaking the $\mathrm{CP}$ symmetry (i.e., $\braket{Q_{\mathit{gluo}}}=0$).

As for $Q_{\mathit{MC}}$, we choose the topological charge~\eqref{eq:clover_charge} measured after $n^\prime_{\mathit{stout}}$ stout smearing steps. This choice allows to adopt the RHMC algorithm in the presence of the topological potential too, as the stout smeared charge is differentiable with respect to the non-stouted links. In our simulations $n^\prime_{\mathit{stout}}$ varies from 2 for the finest lattice spacing up to 10 for the coarsest one.

The path integral expectation value of any observable $\mathcal{O}$ with respect to the original distribution is obtained from expectation values computed in the presence of the bias topological potential through a standard reweighting procedure:
\beq\label{eq:multican_reweight}
\braket{\mathcal{O}} = \frac{\braket{\mathcal{O}e^{V_{\mathit{topo}}(Q_{\mathit{MC}})}}_{\mathit{bias}}}{\braket{e^{V_{\mathit{topo}}(Q_{\mathit{MC}})}}_{\mathit{bias}}}.
\eeq

\section{Results}\label{sec:results}
We computed $\chi$ for several lattice spacings at a temperature $T=\frac{1}{aN_t} \simeq 430~\text{MeV}\simeq 2.8 T_c$ above the transition, adopting both the gluonic and the SP definitions.

The simulation parameters are summarized in Tab.~\ref{tab:simulation_parameters}. The spatial extent of the lattice was chosen so that $L_s \equiv a N_s \sim 1.2$--$1.5$~fm, which is sufficient to contain finite-size effects within our typical statistical error.
\begin{table}[!htb]
\begin{center}
\begin{tabular}{ |c|c|c|c|c|}
\hline
$\beta$ & $a$~[fm] & $\hat{m}_s \cdot 10^{2}$ & $N_s$ & $N_t$ \\
\hline
4.140 & 0.0572 & 2.24 & 32 & 8  \\
4.280 & 0.0458 & 1.81 & 32 & 10 \\
4.385 & 0.0381 & 1.53 & 36 & 12 \\
4.496 & 0.0327 & 1.29 & 48 & 14 \\
4.592 & 0.0286 & 1.09 & 48 & 16 \\
\hline
\end{tabular}
\end{center}
\caption{Summary of simulation parameters. Points refer to temperature $T=1/\left(a N_t\right)\simeq 430~\text{MeV}\simeq 2.8 T_c$. The bare parameters $\beta$ and $\hat{m}_s$ and the lattice spacings have been fixed according to results of Refs.~\cite{Aoki:2009sc, Borsanyi:2010cj, Borsanyi:2013bia}, while $\hat{m}_{l}$ is fixed though $\hat{m}_s/\hat{m}_{l}=m_s/m_l=28.15$.}
\label{tab:simulation_parameters}
\end{table}

\begin{figure}[!htb]
\centering
\includegraphics[scale=0.42]{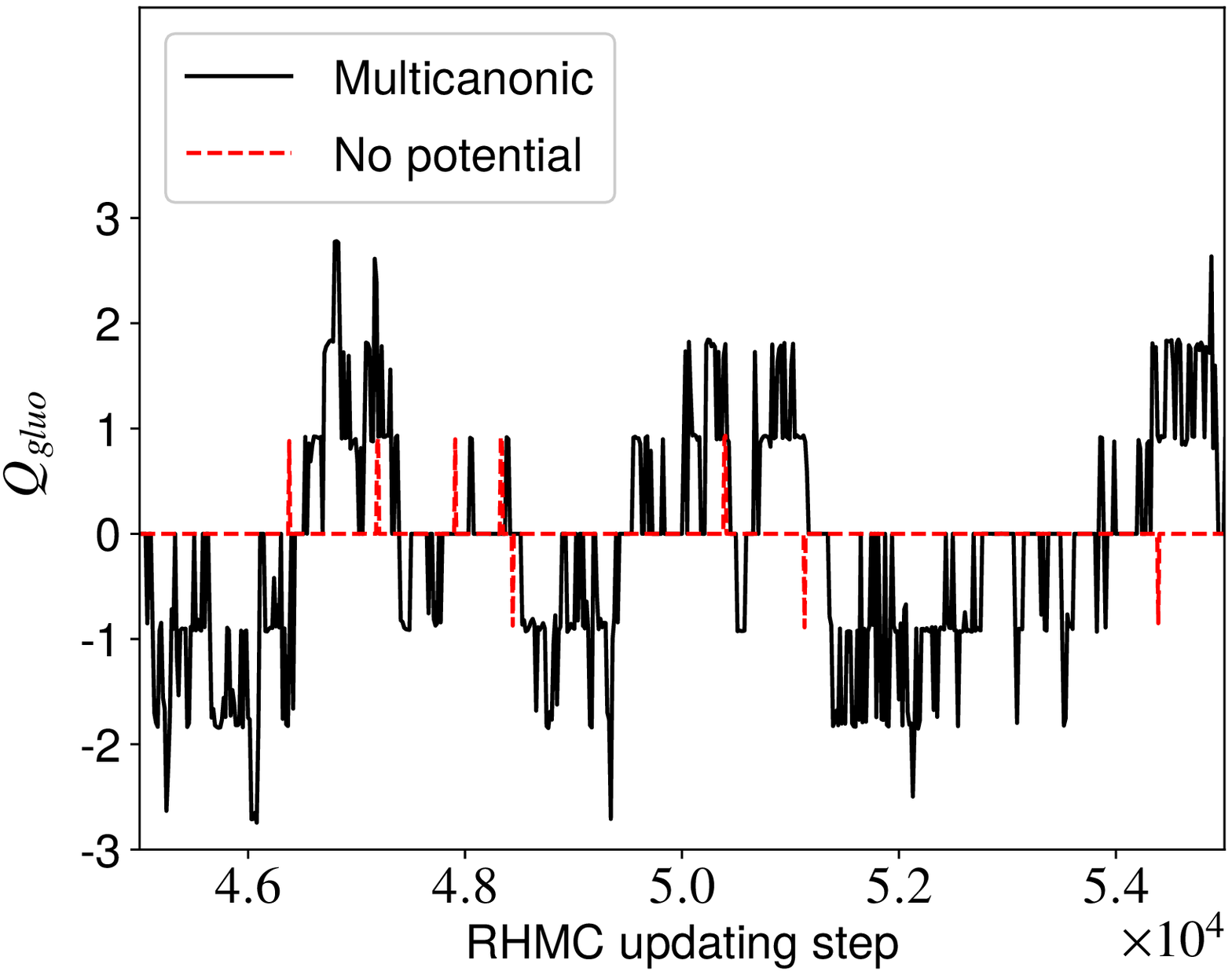}
\hspace{2mm}
\includegraphics[scale=0.42]{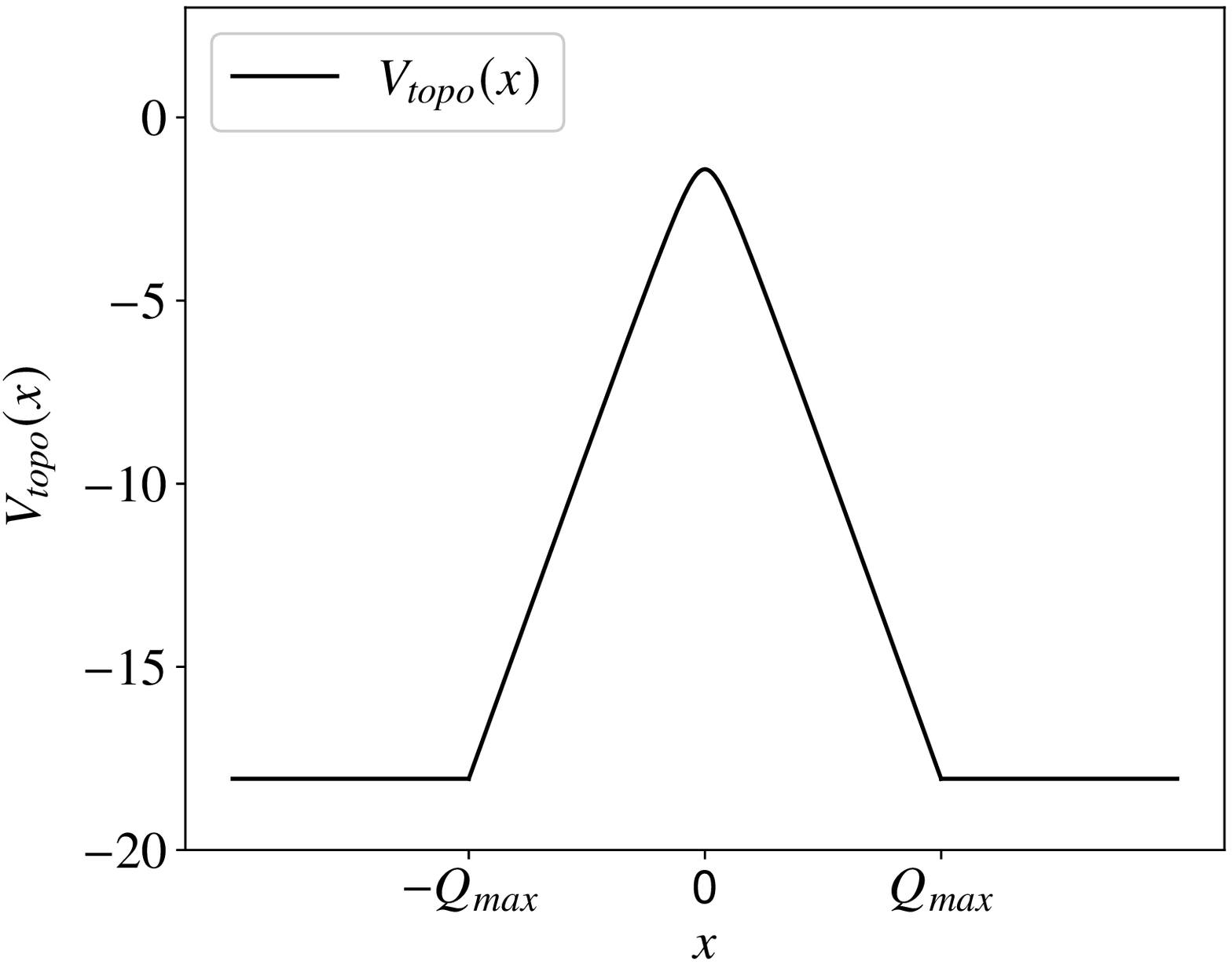}
\caption{Left: Comparison between the Monte Carlo evolution of $Q_{\mathit{gluo}}$ obtained with and without the bias potential for $\beta=4.140$. The time unit used on the horizontal axis is the standard RHMC step. To make the comparison fair, we kept into account that a multicanonical RHMC step is $\sim 60\%$ slower than the standard one. Right: topological potential~\eqref{eq:topo_potential} employed for the $\beta=4.140$ run. In this case $B=6$, $C=2$ and $Q_{\mathit{max}}=3$.}
\label{fig:MC_evolution_Q_multican}
\end{figure}

In Fig.~\ref{fig:MC_evolution_Q_multican} we show an example of the evolution of the topological charge $Q_{\mathit{gluo}}$ obtained in the presence of the bias topological potential for $\beta=4.140$ compared with the corresponding evolution without potential. While fluctuations around $Q_{\mathit{gluo}}=0$ are extremely rare with the standard RHMC, the multicanonic algorithm allows to frequently explore higher-charge topological sectors.

\begin{figure}[!htb]
\centering
\includegraphics[scale=0.42]{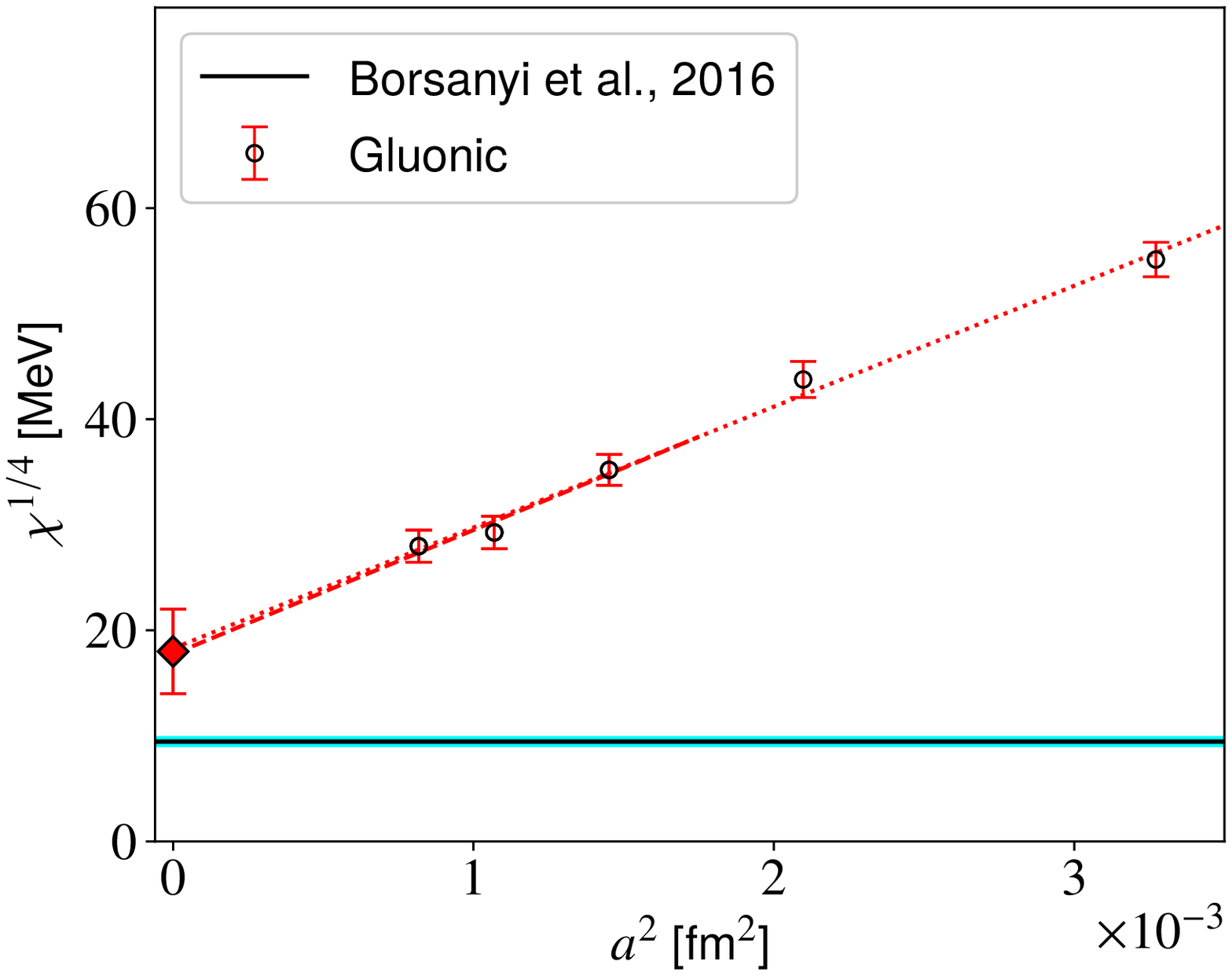}
\includegraphics[scale=0.42]{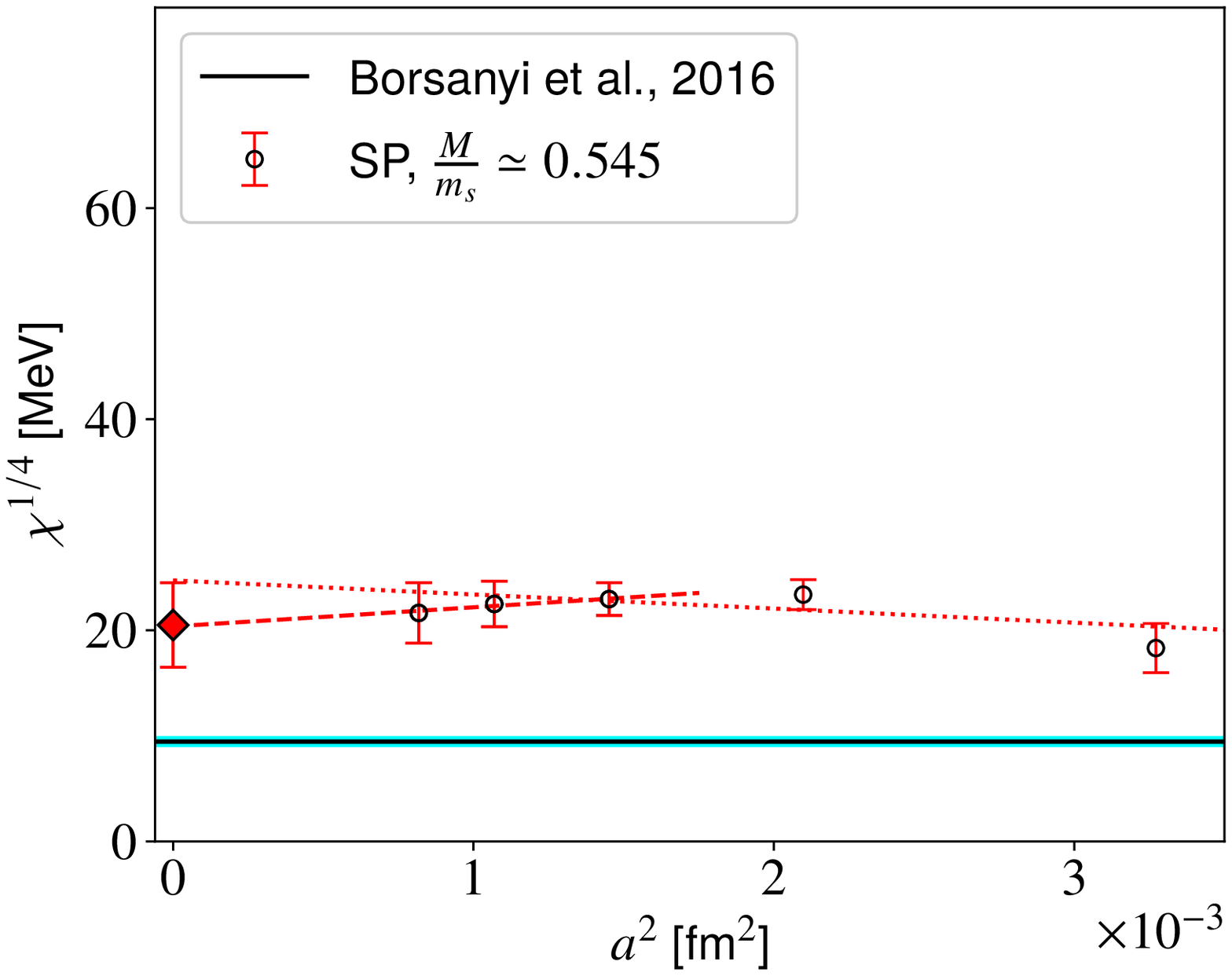}
\includegraphics[scale=0.42]{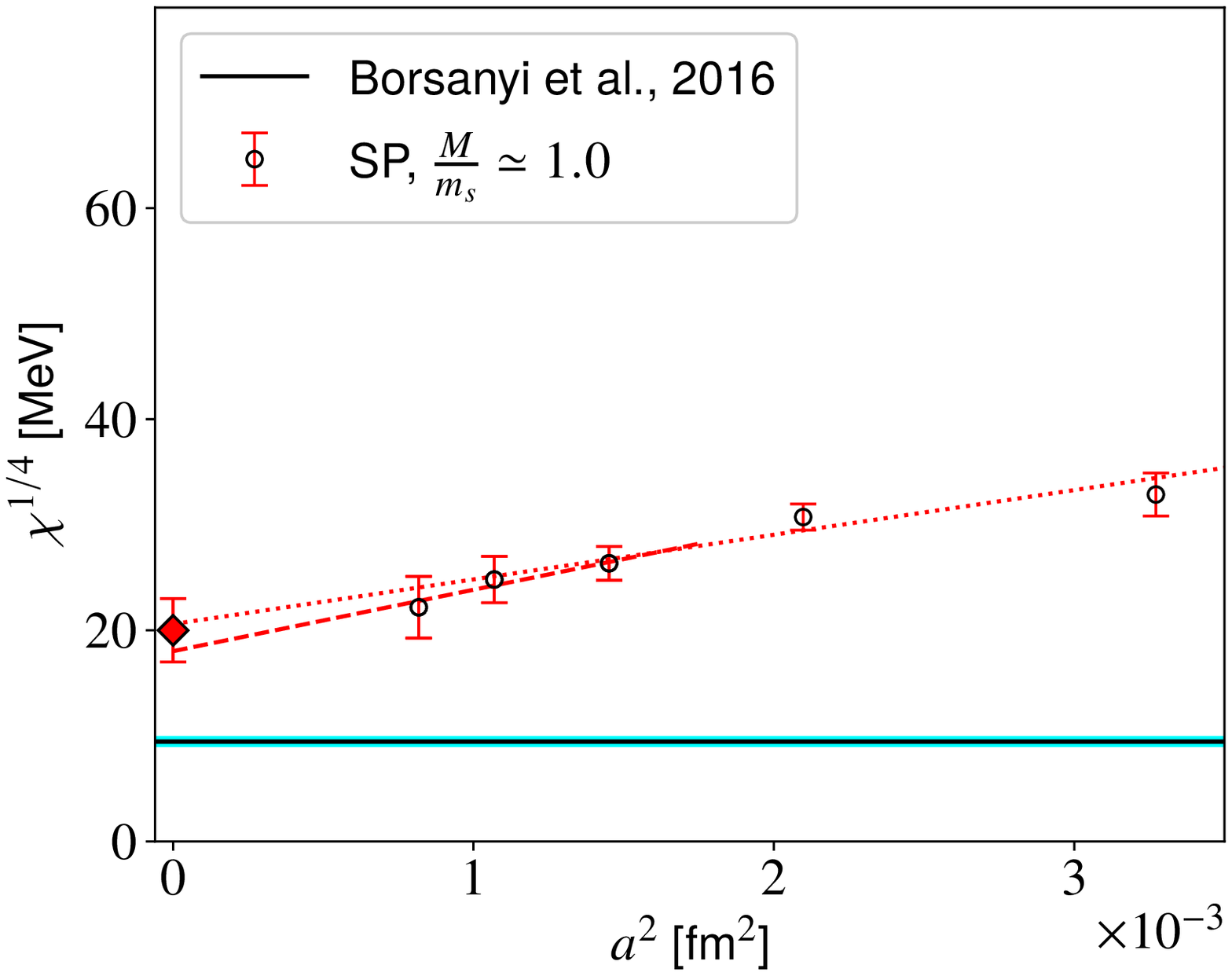}
\caption{Results obtained for the continuum limit of $\chi^{\frac{1}{4}}$ at $T\simeq 430$~MeV using the gluonic and the SP definitions and extrapolating results according to Eq.~\eqref{eq:cont_limit_extrapolation}. Diamond points at $a=0$ represent the continuum extrapolations of the corresponding data sets. The straight line represents $\chi^{\frac{1}{4}}$ obtained at this temperature interpolating results of Ref.~\cite{Borsanyi:2016ksw}.}
\label{fig:continuum_limit}
\end{figure}

In Figs.~\ref{fig:continuum_limit} we show extrapolations towards the continuum limit of finite-lattice-spacing determinations of $\chi_{\mathit{gluo}}$ and $\chi_{\SP}^{(\stag)}$, performed through a best fit of data with a law of the type:
\beq\label{eq:cont_limit_extrapolation}
\chi^{\frac{1}{4}}(a) = \chi_{\mathit{cont}}^{\frac{1}{4}} + c a^2 + o(a^2),
\eeq
where $c$, for SP, depends on the choice of $M/m_s$, while $\chi_{\mathit{cont}}$ is expected to be independent of it.
\FloatBarrier
\FloatBarrier
We extrapolated the continuum limit of the SP determinations of $\chi^{\frac{1}{4}}$ for two values of $M/m_s$ obtained from our multicanonic simulations. In both cases, we obtain consistent results with the continuum limit obtained with the gluonic determination but with more contained lattice artifacts. Moreover, while the magnitude of lattice artifacts does depend on $M/m_s$, we observe no dependence on this parameter of the SP continuum extrapolations, as expected. Thus, systematics related to the choice of the threshold mass are well under control.

We also point out that our result for $\chi^{\frac{1}{4}}$ at this temperature is at odds with the one reported in Ref.~\cite{Borsanyi:2016ksw}, which is about 3 standard deviations far from our determination and a factor of 2 smaller than it, meaning that our result for $\chi$ is more than an order of magnitude larger. The origin of this discrepancy may be due to the different methods that have been adopted to reduce lattice artifacts affecting the continuum scaling of $\chi$, and deserves further investigation to be better clarified.

\section{Conclusions}\label{sec:conclusions}

In this talk we presented preliminary lattice results obtained in $N_f=2+1$ QCD for the topological susceptibility $\chi$ at high temperature from spectral projectors over eigenmodes of the staggered operator. We computed $\chi$ both with the standard gluonic definition and with spectral projectors for 5 different lattice spacings at a temperature $T\simeq 430$~MeV and extrapolated these results towards the continuum limit. The gluonic and the spectral projectors determinations perfectly agree with each other, but spectral projectors suffer from smaller lattice artifacts. In addition, while, we observe no dependence on the choice of the threshold mass $M$ in continuum-extrapolated results, the choice of $M$ affects the magnitude of lattice artifacts, thus this free parameter can be tuned to optimally reduce corrections to the continuum limit.

Comparing our results with the ones obtained in Ref.~\cite{Borsanyi:2016ksw} for the same temperature, we observe a 3-standard-deviation discrepancy. In particular, our determination for $\chi$ is more than an order of magnitude larger. The origin of this discrepancy may be related to the different strategies adopted to reduce the magnitude of lattice artifacts, and it surely deserves to be further investigated. We plan to do so in the near future in a forthcoming work.

Moreover, we also plan to expand this study by adding further temperatures in order to study the behavior of $\chi$ as a function of $T$ above the transition, so that we can compare it with the DIGA prediction as well as with previous determinations in the literature. Note that, being $T=1/(a N_t)$ in the lattice approach, going above a temperature of the order of $\sim 700$~MeV requires lattice spacings of the order of $\sim 0.2$~fm or below for the temporal extents typically employed in currently affordable lattice QCD simulations. For such small lattice spacings the issue of topological freezing becomes extremely relevant, and a strategy to deal with this problem is necessary. In this respect, the adoption of the algorithm recently employed in large-$N$ $SU(N)$ pure-gauge simulations in Ref.~\cite{Bonanno:2020hht} is a possible direction that can be explored in the near future.

\providecommand{\href}[2]{#2}\begingroup\raggedright\endgroup


\begin{thebibliography}{10}
	
	\bibitem{Bonati:2015vqz}
	C.~Bonati, M.~D'Elia, M.~Mariti, G.~Martinelli, M.~Mesiti, F.~Negro et~al.,
	\emph{{Axion phenomenology and $\theta$-dependence from $N_f = 2+1$ lattice
			QCD}}, \href{https://doi.org/10.1007/JHEP03(2016)155}{\emph{JHEP} {\bfseries
			03} (2016) 155} [\href{https://arxiv.org/abs/1512.06746}{{\ttfamily
			1512.06746}}].
	
	\bibitem{Petreczky:2016vrs}
	P.~Petreczky, H.-P. Schadler and S.~Sharma, \emph{{The topological
			susceptibility in finite temperature QCD and axion cosmology}},
	\href{https://doi.org/10.1016/j.physletb.2016.09.063}{\emph{Phys. Lett. B}
		{\bfseries 762} (2016) 498}
	[\href{https://arxiv.org/abs/1606.03145}{{\ttfamily 1606.03145}}].
	
	\bibitem{Frison:2016vuc}
	J.~Frison, R.~Kitano, H.~Matsufuru, S.~Mori and N.~Yamada, \emph{{Topological
			susceptibility at high temperature on the lattice}},
	\href{https://doi.org/10.1007/JHEP09(2016)021}{\emph{JHEP} {\bfseries 09}
		(2016) 021} [\href{https://arxiv.org/abs/1606.07175}{{\ttfamily
			1606.07175}}].
	
	\bibitem{Borsanyi:2016ksw}
	S.~Borsanyi et~al., \emph{{Calculation of the axion mass based on
			high-temperature lattice quantum chromodynamics}},
	\href{https://doi.org/10.1038/nature20115}{\emph{Nature} {\bfseries 539}
		(2016) 69} [\href{https://arxiv.org/abs/1606.07494}{{\ttfamily 1606.07494}}].
	
	\bibitem{Gross:1980br}
	D.~J. Gross, R.~D. Pisarski and L.~G. Yaffe, \emph{{QCD and Instantons at
			Finite Temperature}},
	\href{https://doi.org/10.1103/RevModPhys.53.43}{\emph{Rev. Mod. Phys.}
		{\bfseries 53} (1981) 43}.
	
	\bibitem{Vicari:2008jw}
	E.~Vicari and H.~Panagopoulos, \emph{{Theta dependence of SU(N) gauge theories
			in the presence of a topological term}},
	\href{https://doi.org/10.1016/j.physrep.2008.10.001}{\emph{Phys. Rept.}
		{\bfseries 470} (2009) 93} [\href{https://arxiv.org/abs/0803.1593}{{\ttfamily
			0803.1593}}].
	
	\bibitem{Schaefer:2010hu}
	{\scshape ALPHA} collaboration, S.~Schaefer, R.~Sommer and F.~Virotta,
	\emph{{Critical slowing down and error analysis in lattice QCD simulations}},
	\href{https://doi.org/10.1016/j.nuclphysb.2010.11.020}{\emph{Nucl. Phys. B}
		{\bfseries 845} (2011) 93} [\href{https://arxiv.org/abs/1009.5228}{{\ttfamily
			1009.5228}}].
	
	\bibitem{Bonati:2017woi}
	C.~Bonati and M.~D'Elia, \emph{{Topological critical slowing down: variations
			on a toy model}},
	\href{https://doi.org/10.1103/PhysRevE.98.013308}{\emph{Phys. Rev. E}
		{\bfseries 98} (2018) 013308}
	[\href{https://arxiv.org/abs/1709.10034}{{\ttfamily 1709.10034}}].
	
	\bibitem{Luscher:2004fu}
	M.~L{\"u}scher, \emph{{Topological effects in QCD and the problem of short
			distance singularities}},
	\href{https://doi.org/10.1016/j.physletb.2004.04.076}{\emph{Phys. Lett. B}
		{\bfseries 593} (2004) 296}
	[\href{https://arxiv.org/abs/hep-th/0404034}{{\ttfamily hep-th/0404034}}].
	
	\bibitem{Giusti:2008vb}
	L.~Giusti and M.~L{\"u}scher, \emph{{Chiral symmetry breaking and the
			Banks-Casher relation in lattice QCD with Wilson quarks}},
	\href{https://doi.org/10.1088/1126-6708/2009/03/013}{\emph{JHEP} {\bfseries
			03} (2009) 013} [\href{https://arxiv.org/abs/0812.3638}{{\ttfamily
			0812.3638}}].
	
	\bibitem{Luscher:2010ik}
	M.~L{\"u}scher and F.~Palombi, \emph{{Universality of the topological
			susceptibility in the $SU(3)$ gauge theory}},
	\href{https://doi.org/10.1007/JHEP09(2010)110}{\emph{JHEP} {\bfseries 09}
		(2010) 110} [\href{https://arxiv.org/abs/1008.0732}{{\ttfamily 1008.0732}}].
	
	\bibitem{Cichy:2015jra}
	{\scshape ETM} collaboration, K.~Cichy, E.~Garcia-Ramos, K.~Jansen, K.~Ottnad
	and C.~Urbach, \emph{{Non-perturbative Test of the Witten-Veneziano Formula
			from Lattice QCD}},
	\href{https://doi.org/10.1007/JHEP09(2015)020}{\emph{JHEP} {\bfseries 09}
		(2015) 020} [\href{https://arxiv.org/abs/1504.07954}{{\ttfamily
			1504.07954}}].
	
	\bibitem{Alexandrou:2017bzk}
	C.~Alexandrou, A.~Athenodorou, K.~Cichy, M.~Constantinou, D.~P. Horkel,
	K.~Jansen et~al., \emph{{Topological susceptibility from twisted mass
			fermions using spectral projectors and the gradient flow}},
	\href{https://doi.org/10.1103/PhysRevD.97.074503}{\emph{Phys. Rev. D}
		{\bfseries 97} (2018) 074503}
	[\href{https://arxiv.org/abs/1709.06596}{{\ttfamily 1709.06596}}].
	
	\bibitem{Bonanno:2019xhg}
	C.~Bonanno, G.~Clemente, M.~D'Elia and F.~Sanfilippo, \emph{{Topology via
			spectral projectors with staggered fermions}},
	\href{https://doi.org/10.1007/JHEP10(2019)187}{\emph{JHEP} {\bfseries 10}
		(2019) 187} [\href{https://arxiv.org/abs/1908.11832}{{\ttfamily
			1908.11832}}].
	
	\bibitem{Jahn:2018dke}
	P.~T. Jahn, G.~D. Moore and D.~Robaina, \emph{{$\chi_{\textrm{top}}(T \gg
			T_{\textrm{c}})$ in pure-glue QCD through reweighting}},
	\href{https://doi.org/10.1103/PhysRevD.98.054512}{\emph{Phys. Rev. D}
		{\bfseries 98} (2018) 054512}
	[\href{https://arxiv.org/abs/1806.01162}{{\ttfamily 1806.01162}}].
	
	\bibitem{Bonati:2018blm}
	C.~Bonati, M.~D'Elia, G.~Martinelli, F.~Negro, F.~Sanfilippo and A.~Todaro,
	\emph{{Topology in full QCD at high temperature: a multicanonical approach}},
	\href{https://doi.org/10.1007/JHEP11(2018)170}{\emph{JHEP} {\bfseries 11}
		(2018) 170} [\href{https://arxiv.org/abs/1807.07954}{{\ttfamily
			1807.07954}}].
	
	\bibitem{Morningstar:2003gk}
	C.~Morningstar and M.~J. Peardon, \emph{{Analytic smearing of SU(3) link
			variables in lattice QCD}},
	\href{https://doi.org/10.1103/PhysRevD.69.054501}{\emph{Phys. Rev. D}
		{\bfseries 69} (2004) 054501}
	[\href{https://arxiv.org/abs/hep-lat/0311018}{{\ttfamily hep-lat/0311018}}].
	
	\bibitem{Clark:2006fx}
	M.~A. Clark and A.~D. Kennedy, \emph{{Accelerating dynamical fermion
			computations using the rational hybrid Monte Carlo (RHMC) algorithm with
			multiple pseudofermion fields}},
	\href{https://doi.org/10.1103/PhysRevLett.98.051601}{\emph{Phys. Rev. Lett.}
		{\bfseries 98} (2007) 051601}
	[\href{https://arxiv.org/abs/hep-lat/0608015}{{\ttfamily hep-lat/0608015}}].
	
	\bibitem{Clark:2006wp}
	M.~A. Clark and A.~D. Kennedy, \emph{{Accelerating Staggered Fermion Dynamics
			with the Rational Hybrid Monte Carlo (RHMC) Algorithm}},
	\href{https://doi.org/10.1103/PhysRevD.75.011502}{\emph{Phys. Rev. D}
		{\bfseries 75} (2007) 011502}
	[\href{https://arxiv.org/abs/hep-lat/0610047}{{\ttfamily hep-lat/0610047}}].
	
	\bibitem{Aoki:2009sc}
	Y.~Aoki, S.~Borsanyi, S.~Durr, Z.~Fodor, S.~D. Katz, S.~Krieg et~al.,
	\emph{{The QCD transition temperature: results with physical masses in the
			continuum limit II.}},
	\href{https://doi.org/10.1088/1126-6708/2009/06/088}{\emph{JHEP} {\bfseries
			06} (2009) 088} [\href{https://arxiv.org/abs/0903.4155}{{\ttfamily
			0903.4155}}].
	
	\bibitem{Borsanyi:2010cj}
	S.~Borsanyi, G.~Endrodi, Z.~Fodor, A.~Jakovac, S.~D. Katz, S.~Krieg et~al.,
	\emph{{The QCD equation of state with dynamical quarks}},
	\href{https://doi.org/10.1007/JHEP11(2010)077}{\emph{JHEP} {\bfseries 11}
		(2010) 077} [\href{https://arxiv.org/abs/1007.2580}{{\ttfamily 1007.2580}}].
	
	\bibitem{Borsanyi:2013bia}
	S.~Borsanyi, Z.~Fodor, C.~Hoelbling, S.~D. Katz, S.~Krieg and K.~K. Szabo,
	\emph{{Full result for the QCD equation of state with 2+1 flavors}},
	\href{https://doi.org/10.1016/j.physletb.2014.01.007}{\emph{Phys. Lett. B}
		{\bfseries 730} (2014) 99} [\href{https://arxiv.org/abs/1309.5258}{{\ttfamily
			1309.5258}}].
	
	\bibitem{Campostrini:1988cy}
	M.~Campostrini, A.~Di~Giacomo and H.~Panagopoulos, \emph{{The Topological
			Susceptibility on the Lattice}},
	\href{https://doi.org/10.1016/0370-2693(88)90526-6}{\emph{Phys. Lett. B}
		{\bfseries 212} (1988) 206}.
	
	\bibitem{DiVecchia:1981aev}
	P.~Di~Vecchia, K.~Fabricius, G.~C. Rossi and G.~Veneziano, \emph{{Preliminary
			Evidence for $U(1)_A$ Breaking in QCD from Lattice Calculations}},
	\href{https://doi.org/10.1016/0550-3213(81)90432-6}{\emph{Nucl. Phys. B}
		{\bfseries 192} (1981) 392}.
	
	\bibitem{DiVecchia:1981hh}
	P.~Di~Vecchia, K.~Fabricius, G.~Rossi and G.~Veneziano, \emph{{Numerical Checks
			of the Lattice Definition Independence of Topological Charge Fluctuations}},
	\href{https://doi.org/10.1016/0370-2693(82)91203-5}{\emph{Phys. Lett. B}
		{\bfseries 108} (1982) 323}.
	
	\bibitem{Berg:1981nw}
	B.~Berg, \emph{{Dislocations and Topological Background in the Lattice $O(3)$
			$\sigma$ Model}},
	\href{https://doi.org/10.1016/0370-2693(81)90518-9}{\emph{Phys. Lett. B}
		{\bfseries 104} (1981) 475}.
	
	\bibitem{Iwasaki:1983bv}
	Y.~Iwasaki and T.~Yoshie, \emph{{Instantons and Topological Charge in Lattice
			Gauge Theory}},
	\href{https://doi.org/10.1016/0370-2693(83)91111-5}{\emph{Phys. Lett. B}
		{\bfseries 131} (1983) 159}.
	
	\bibitem{Itoh:1984pr}
	S.~Itoh, Y.~Iwasaki and T.~Yoshie, \emph{{Stability of Instantons on the
			Lattice and the Renormalized Trajectory}},
	\href{https://doi.org/10.1016/0370-2693(84)90609-9}{\emph{Phys. Lett. B}
		{\bfseries 147} (1984) 141}.
	
	\bibitem{Teper:1985rb}
	M.~Teper, \emph{{Instantons in the Quantized $SU(2)$ Vacuum: A Lattice Monte
			Carlo Investigation}},
	\href{https://doi.org/10.1016/0370-2693(85)90939-6}{\emph{Phys. Lett. B}
		{\bfseries 162} (1985) 357}.
	
	\bibitem{Ilgenfritz:1985dz}
	E.-M. Ilgenfritz, M.~Laursen, G.~Schierholz, M.~M{\"u}ller-Preussker and
	H.~Schiller, \emph{{First Evidence for the Existence of Instantons in the
			Quantized $SU(2)$ Lattice Vacuum}},
	\href{https://doi.org/10.1016/0550-3213(86)90265-8}{\emph{Nucl. Phys. B}
		{\bfseries 268} (1986) 693}.
	
	\bibitem{Campostrini:1989dh}
	M.~Campostrini, A.~Di~Giacomo, H.~Panagopoulos and E.~Vicari,
	\emph{{Topological Charge, Renormalization and Cooling on the Lattice}},
	\href{https://doi.org/10.1016/0550-3213(90)90077-Q}{\emph{Nucl. Phys. B}
		{\bfseries 329} (1990) 683}.
	
	\bibitem{Alles:2000sc}
	B.~Alles, L.~Cosmai, M.~D'Elia and A.~Papa, \emph{{Topology in $2D$ $CP^{N-1}$
			models on the lattice: A Critical comparison of different cooling
			techniques}}, \href{https://doi.org/10.1103/PhysRevD.62.094507}{\emph{Phys.
			Rev. D} {\bfseries 62} (2000) 094507}
	[\href{https://arxiv.org/abs/hep-lat/0001027}{{\ttfamily hep-lat/0001027}}].
	
	\bibitem{Luscher:2009eq}
	M.~L{\"u}scher, \emph{{Trivializing maps, the Wilson flow and the HMC
			algorithm}}, \href{https://doi.org/10.1007/s00220-009-0953-7}{\emph{Commun.
			Math. Phys.} {\bfseries 293} (2010) 899}
	[\href{https://arxiv.org/abs/0907.5491}{{\ttfamily 0907.5491}}].
	
	\bibitem{Luscher:2010iy}
	M.~L{\"u}scher, \emph{{Properties and uses of the Wilson flow in lattice QCD}},
	\href{https://doi.org/10.1007/JHEP08(2010)071,
		10.1007/JHEP03(2014)092}{\emph{JHEP} {\bfseries 08} (2010) 071}
	[\href{https://arxiv.org/abs/1006.4518}{{\ttfamily 1006.4518}}].
	
	\bibitem{Bonati:2014tqa}
	C.~Bonati and M.~D'Elia, \emph{{Comparison of the gradient flow with cooling in
			$SU(3)$ pure gauge theory}},
	\href{https://doi.org/10.1103/PhysRevD.89.105005}{\emph{Phys. Rev. D}
		{\bfseries D89} (2014) 105005}
	[\href{https://arxiv.org/abs/1401.2441}{{\ttfamily 1401.2441}}].
	
	\bibitem{Alexandrou:2015yba}
	C.~Alexandrou, A.~Athenodorou and K.~Jansen, \emph{{Topological charge using
			cooling and the gradient flow}},
	\href{https://doi.org/10.1103/PhysRevD.92.125014}{\emph{Phys. Rev. D}
		{\bfseries 92} (2015) 125014}
	[\href{https://arxiv.org/abs/1509.04259}{{\ttfamily 1509.04259}}].
	
	\bibitem{DelDebbio:2002xa}
	L.~Del~Debbio, H.~Panagopoulos and E.~Vicari, \emph{{theta dependence of
			$SU(N)$ gauge theories}},
	\href{https://doi.org/10.1088/1126-6708/2002/08/044}{\emph{JHEP} {\bfseries
			08} (2002) 044} [\href{https://arxiv.org/abs/hep-th/0204125}{{\ttfamily
			hep-th/0204125}}].
	
	\bibitem{Bonanno:2020hht}
	C.~Bonanno, C.~Bonati and M.~D'Elia, \emph{{Large-$N$ $SU(N)$ Yang-Mills
			theories with milder topological freezing}},
	\href{https://doi.org/10.1007/JHEP03(2021)111}{\emph{JHEP} {\bfseries 03}
		(2021) 111} [\href{https://arxiv.org/abs/2012.14000}{{\ttfamily
			2012.14000}}].
	
\end{thebibliography}
\end{document}